\documentclass[twocolumn,english,aps,showpacs,prl,superscriptaddress,preprintnumbers,floatfix,footinbib]{revtex4}

\begin{document}

\preprint{DAPNIA-04-37,ADP-04-03/T582}

\title{Quark structure and nuclear effective forces}

\author{P.A.M. Guichon}

\affiliation{SPhN-DAPNIA, CEA Saclay, F91191 Gif sur Yvette, France}

\author{A.W. Thomas}

\affiliation{Special Research Centre for the Subatomic Structure of Matter, 
University of Adelaide, SA 5005, Australia}

\begin{abstract}
We formulate the quark meson coupling model as a many-body effective
Hamiltonian. This leads naturally to the appearance of many-body forces.
We investigate the zero range limit of the model and compare its
Hartree-Fock Hamiltonian to that corresponding to the Skyrme effective
force. By fixing the three parameters of the model to reproduce the binding and
symmetry energy of nuclear matter, we find that it
allows a very satisfactory interpretation of the Skyrme force.
\end{abstract}
\maketitle
The notion that quark degrees of freedom may play a role in low energy
nuclear physics is largely unappreciated.
The main reason is probably that the many
body formulation of nuclear physics based on point like nucleons interacting
through effective forces has proven quite successful. In this paper we take
a radically different point of view, arguing that the nuclear
effective force itself is a direct manifestation of the quark structure
of the nucleon. To this end we formulate the quark meson coupling
(QMC) model of the nucleus as a many body problem. This allows us to
take the limit corresponding
to a zero range force which can be compared to the Skyrme force\cite{Skyrme55}. 

In the quark meson coupling model \cite{Guichon88,Guichon96} the
essential step is to solve for the quark structure of the nucleon under
the influence of the nuclear environment. For this one considers
that, in a time averaged sense, a nucleus can be described as a collection
of non overlapping quark bags representing the nucleons. (More recently
the same ideas have been extended to a confined version of the 
NJL model~\cite{Bentz:2001vc}.) 
The interactions
of the quarks with the nuclear medium is represented by the exchange
of mesons between the quarks of {\it different} nucleons, with coupling
constants treated as free parameters. As explained in Ref.\cite{key-11}
the \( \sigma  \) field which is used here is not the chiral partner
of the pion and the quark-\( \sigma  \) coupling does not break chiral
invariance.

As in our previous work \cite{Guichon96}
the scalar field is denoted \( \sigma (\vec{r}) \), while 
\( \omega (\vec{r}) \) is the time component of the vector
field and both are taken to be time independent. In the nuclear ground
state, or for low energy excited states, the time dependence of the fields 
is driven by the Fermi motion
of the nucleons, so the typical frequencies are of the order of the
Fermi energy, which can be neglected with respect to the high frequencies
of the confined quark fields. Moreover, the space components of the
\( \omega  \) field have their source in the velocity density of the
nucleons, which is not a coherent quantity -- in contrast to 
the nucleon density \( \rho (\vec{r}) \),
the source of the time component. Therefore, when we solve for the nucleon
structure under the influence of the medium, the dominant effect comes
{}from the time component~\footnote{Of course in the free 
$NN$ interaction one must treat all components of the meson field 
on an equal footing.}.

Each bag is moving in the classical fields,
\( \sigma (\vec{r}),\omega (\vec{r}) \), 
to which the quarks are coupled. In the spirit of the Born-Oppenheimer
approximation we solve the equations of motion of the quarks in a
given bag for a fixed classical position, \( \vec{R}(t) \), of its
center. For this we use the known Lorentz character of the $\sigma$ and
$\omega$ fields to transform to the instantaneous rest frame of the bag, 
where the static spherical cavity approximation 
is most appropriate. 
We then expand the fields around 
their values at the center of the bag, truncating the expansion at
first order. The coupling of the quark to the constant part of the
fields is solved exactly, because it amounts to a shift in the quark
mass and energy. The remainder is treated as a perturbation. The rest frame
energy-momentum is then transformed back to the nuclear rest frame
with proper account of Thomas spin precession. Keeping only terms
which are quadratic in the nucleon velocity, as we do systematically
throughout this work, we find the following expression for the
classical energy of a nucleon with position-momentum 
\( (\vec{R},\, \vec{P}) \)~\cite{Guichon96}:
\begin{eqnarray}
E_{N}(\vec{R}) & = & \frac{\vec{P}^{2}}{2M^{*}(\vec{R})}+M^{*}(\vec{R})+
g_{\omega }\omega (\vec{R})+V_{so} \, ,  
\label{Eq-QMC1} 
\end{eqnarray}
with \( g_{\omega } \)the \( \omega  \)-nucleon coupling constant. 
The spin-orbit interaction, \( V_{so} \), is defined below.
To get the dynamical mass \( M^{*}(\vec{R}) \)
one has to solve the bag equations in the field \( \sigma (\vec{R}) \).
For our purpose it is sufficient to know that it is well approximated
by the expression
\begin{equation}
\label{Eq-QMC4}
M^{*}(\vec{R})=M-g_{\sigma }\sigma (\vec{R})+
\frac{d}{2}\left( g_{\sigma }\sigma (\vec{R})\right) ^{2} \, ,  
\end{equation}
where \( (M,\, g_{\sigma }) \) are the mass and the \( \sigma  \)-nucleon
coupling constant for the free nucleon and \( d=0.22R_{B} \) -- with
$R_{B}$ the bag radius. The last term, which represents the response of the
nucleon to the applied scalar field (the ''scalar polarizability''),   
is an essential element of the QMC
model. From
our numerical studies we know that the approximation (\ref{Eq-QMC4}) 
is quite accurate 
up to \( g_{\sigma }\sigma =400 \) MeV, which should be sufficient
for our purposes. As a check we have also tried a cubic polynomial 
but found no significant changes in the results. 

The energy (\ref{Eq-QMC1}) is the energy of one particular nucleon
moving classically in the nuclear meson fields. Since, by hypothesis, 
the bags do not overlap, the total energy of the system is the sum
of the energy of each of the nucleons plus the energy carried by
the fields. As the latter are static we can write:
\begin{eqnarray}
&&E_{tot}=\sum _{i}E_{N}(\vec{R}_{i})+E_{mes},  \\
&&E_{mes}=\frac{1}{2}\int d\vec{r} \left[
\left( {\nabla }\sigma \right) ^{2} + 
m_{\sigma }^{2}\sigma ^{2}-\left( {\nabla }\omega 
\right) ^{2}-m_{\omega }^{2}\omega ^{2} \right],\nonumber\\ 
\end{eqnarray}
with \( m_{\sigma },\, m_{\omega } \) the masses of the mesons. 

To simplify the expression for \( E_{N}(\vec{R}) \) we estimate the
quantity \( g_{\sigma }\sigma  \) using the field equations 
\( \delta E_{tot}/\delta \sigma (\vec{r})=0 \).
Neglecting the velocity dependent terms, setting 
\( M^{*}\approx M-g_{\sigma }\sigma  \)
and neglecting \( ( {\nabla }\sigma )^{2} \)
with respect to \( m_{\sigma }^{2}\sigma ^{2} \), we find  
\( g_{\sigma }\sigma (\vec{r})\sim G_{\sigma }\rho ^{cl}(\vec{r}) \),  
with \( G_{\sigma }=g_{\sigma }^{2}/m_{\sigma }^{2} \) and the classical density
is defined as $\rho ^{cl}(\vec{r})=\sum _{i}\delta (\vec{r}-\vec{R}_{i})$.

{}From our previous studies in the Hartree approximation, 
\( G_{\sigma }\sim 10{\rm fm}^{2} \), which yields \( g_{\sigma }\sigma \simeq
300 \) MeV at nuclear matter density. 
Making a quadratic expansion of \( 1/M^{*} \) in 
powers of \( g_{\sigma }\sigma  \)
and keeping the leading terms~\footnote{We use the natural hierarchy, 
$1 >> g_{\sigma}\sigma d/2 \sim 0.17 >> \\ <\vec{P}^2 / 2 M^2> \sim 0.025$, to 
systematize the expansion.} we obtain: 
\begin{eqnarray}
\frac{\vec{P}^{2}}{2M^{*}(\vec{R})}+
M^{*}(\vec{R})\approx M+\frac{\vec{P}^{2}}{2M} \, \, - & \nonumber \\
g_{\sigma }\sigma (\vec{R})\left( 1-\frac{d}{2}g_{\sigma }\sigma
(\vec{R})\right) \left( 1-\frac{\vec{P}^{2}}{2M^{2}}\right)\,   &  .  
\end{eqnarray}
If we define the scalar density as $\rho ^{cl}_{s}(\vec{r})=
\sum _{i}(1-{\vec{P}^{2}_{i}}/2M^{2}) \delta (\vec{r}-\vec{R}_{i})$, we can write the total energy in the form 
\begin{eqnarray}
&  & E_{tot}=E_{mes+}\sum _{i}\left( M+
\frac{\vec{P}^{2}_{i}}{2M}+V_{so}(i)\right) -\nonumber \\
&  & \int d\vec{r}\, \rho ^{cl}_{s}\, \left( 
g_{\sigma }\sigma -\frac{d}{2}(g_{\sigma }\sigma )^{2}\right) + 
\int d\vec{r}\, \rho ^{cl}\, g_{\omega }\omega \, , 
\label{Eq-QMC10} 
\end{eqnarray}
which will be our starting point for the many body formulation of
the QMC model. We use the equations for the mesons,
\( \delta E_{tot}/\delta \sigma (\vec{r})=
\delta E_{tot}/\delta \omega (\vec{r})=0 \), 
to eliminate the meson fields from the energy, leaving a system whose
dynamics depends only on the nucleon coordinates. This is possible
because the conjugate momenta of the static meson fields vanish identically.
{}From Eq.(\ref{Eq-QMC10}) we write the equations for the meson fields
in the following forms:
\begin{eqnarray}
g_{\sigma }\sigma  & = & G_{\sigma }\rho ^{cl}_{s}(1-d\, g_{\sigma }\sigma )+
\nabla ^{2}g_{\sigma }\sigma /m_{\sigma }^{2}\, ,\label{Eq-QMC13} \\
g_{\omega }\omega  & = & G_{\omega }\rho ^{cl}+
\nabla ^{2}g_{\omega }\omega /m_{\omega }^{2}\, ,\label{Eq-QMC14} 
\end{eqnarray}
where we have defined \( G_{\omega }=g_{\omega }^{2}/m_{\omega }^{2}. \)
On the RHS of Eqs.~(\ref{Eq-QMC13},\ref{Eq-QMC14}) we have neglected
the contribution of the functional derivative acting on the spin-orbit
term in (\ref{Eq-QMC10}). This is because the latter was obtained
as a first order perturbation and one can check that the resulting
error in the final Hamiltonian is of higher order. If we insert the
solutions \( \omega _{sol}(\vec{r}),\, \sigma _{sol}(\vec{r}) \)
of Eqs.~(\ref{Eq-QMC13},\ref{Eq-QMC14}) in the expression (\ref{Eq-QMC10})
{}for the energy we get, after some algebra, and omitting the irrelevant
constant mass term:
\begin{eqnarray}
E_{tot} & = & \sum _{i}\left( 
\frac{\vec{P}^{2}_{i}}{2M}+V_{so}(i)\right) \nonumber \\
& - & \frac{1}{2}\int d\vec{r}\, \rho ^{cl}_{s}\, 
g_{\sigma }\sigma _{sol}+\frac{1}{2}\int d\vec{r}\, \rho ^{cl}\, 
g_{\omega }\omega _{sol} \, .  
\label{Eq-QMC15} 
\end{eqnarray}

We do not attempt to use exact solutions of Eqs.(\ref{Eq-QMC13},\ref{Eq-QMC14})
as this would lead to an intricate many body problem that would be
difficult to compare with standard nuclear physics approaches. Instead
we first remark that, roughly speaking, the meson fields should follow
the matter density. Therefore the typical scale for the \( \nabla  \)
operator acting on \( \sigma  \) or \( \omega  \) is the thickness
of the nuclear surface, that is about \( 1{\rm fm}. \) 
In so far as \( 1{\rm fm}^{-2}\ll (m_{\sigma }^{2},m_{\omega }^{2}) \), 
which looks reasonable, we can consider the terms 
\( \nabla ^{2}g_{\sigma }\sigma /m_{\sigma }^{2} \)
and \( \nabla ^{2}g_{\omega }\omega /m_{\omega }^{2} \) as perturbations
and, \emph{in these terms}, replace \( \sigma  \) and \( \omega  \)
by their first order approximation, that is:
\( g_{\sigma }\sigma \approx G_{\sigma }\rho _{s}^{cl} \)
and \( g_{\omega }\omega \approx G_{\omega }\rho ^{cl} \). The next
step in solving for the \( \sigma  \) field is to
solve Eq.~(\ref{Eq-QMC13}) iteratively , starting from the lowest order
approximation, \( g_{\sigma }\sigma =G_{\sigma }\rho ^{cl}_{s}. \)
When inserted in Eq.~(\ref{Eq-QMC15}) this series will generate N-body
forces with convergence controlled by the parameter 
\( d\, g_{\sigma }\sigma \simeq 0.33 \) , 
according to our estimate. To simplify further we shall neglect the
small difference between \( \rho ^{cl}_{s} \) and \( \rho ^{cl} \)
except in the leading term, the one which generates the 2-body force.
These approximations will not be difficult to improve but, as this
leads inevitably to an effective interaction which is more complicated
than the simple Skyrme force, we postpone this to future investigations.
In summary the expressions we shall use for the field solutions are
: 
\begin{eqnarray}
& & g_{\sigma }\sigma _{sol}(\vec{r}) = \frac{G_{\sigma }}{m_{\sigma }^{2}}\nabla ^{2}\rho ^{cl}
 +G_{\sigma }\rho ^{cl}_{s} +
 \sum _{k\geq 1}(-d)^{k}\left( G_{\sigma }\rho ^{cl}\right) ^{k+1},\nonumber \label{Eq-QMC19} \\
& & \\
& & g_{\omega }\omega _{sol}(\vec{r}) = \frac{G_{\omega }}{m_{\omega
 }^{2}}\nabla ^{2}\rho ^{cl}+G_{\omega }\rho ^{cl} \, .  
\end{eqnarray}

The rest of the derivation amounts to substituting 
\( g_{\omega }\omega _{sol} \)
and \( g_{\sigma }\sigma _{sol} \) into Eq.~(\ref{Eq-QMC15}) for  
the energy. As usual the density and the scalar density to some
power contain infinite terms corresponding to the self-interaction
of the nucleon. Formally these can be incorporated
into a redefinition of the mass and coupling constants of the free nucleon.
Since our model is devised to describe the modification of the nucleon
by the medium rather than the nucleon itself we simply remove them.
This amounts to the replacements 
\( 
( \sum _{i}\delta (\vec{r}-\vec{R}_{i}) )^{2}
\rightarrow \sum _{i\neq j}\delta 
(\vec{r}-\vec{R}_{i})\delta (\vec{r}-\vec{R}_{j}) 
\), which
leads to the following many-body Hamiltonian, essentially equivalent
to the QMC model: 
\begin{eqnarray}
 &  & H_{QMC}=\sum _{i}\left( 
\frac{\vec{P}^{2}_{i}}{2M}+V_{so}(i)\right) +\frac{G_{\sigma }}{2}\sum
_{i\neq j}\frac{\vec{P}^{2}_{i}}{M^{2}}\delta (\vec{R}_{ij})\nonumber \\
&  & +\frac{G_{\omega }}{2}\sum _{i\neq j}\left( \delta (\vec{R}_{ij})+\frac{1}{m_{\omega }^{2}}\nabla _{i}^{2}\delta (\vec{R}_{ij})\right) \nonumber \\
&  & -\frac{G_{\sigma }}{2}\sum _{i\neq j}\left( \delta (\vec{R}_{ij})+\frac{1}{m_{\sigma }^{2}}\nabla _{i}^{2}\delta (\vec{R}_{ij})\right) \nonumber \\
&  & +\frac{dG_{\sigma }^{2}}{2}\sum _{i\neq j\neq k}\delta ^{2}(ijk)-
\frac{d^{2}G_{\sigma }^{3}}{2}\sum _{i\neq j\neq k\neq l}\delta ^{3}(ijkl)
\, . 
\label{Eq-QMC21} 
\end{eqnarray}
Here \( \vec{R}_{ij}=\vec{R}_{i}-\vec{R}_{j} \) and \( \nabla _{i} \)
is the gradient with respect to \( \vec{R}_{i} \) acting on the delta
function. In Eq.(\ref{Eq-QMC21}) we have dropped the contact interactions
involving more than 4-bodies, because their matrix elements vanish
for antisymmetrized states. To shorten the equations we used the notation
\( \delta ^{2}(ijk) \) for \( \delta (\vec{R}_{ij})\delta (\vec{R}_{jk}) \)
and analogously for \( \delta ^{3}(ijkl) \) . For the spin orbit
interaction we start from our previous result~\cite{Guichon96} :
\begin{equation}
\label{Eq-QMC22}
\sum _{i}V_{so}(i)=\sum _{i}\frac{1}{4M^{*2}(\vec{R}_{i})}\vec{P}_{i}
\times {\nabla }_{_{i}}W(\vec{R}_{i}).\vec{\sigma }_{i}\, ,
\end{equation}
where 
\( W(\vec{R}_{i})=M^{*}(\vec{R}_{i})+g_{\omega }\omega 
(\vec{R}_{i})(1-2\mu _{s}) \),
\( \mu _{s}=0.9 \) is the isoscalar magnetic moment and \( \vec{\sigma }_{i} \)
are the Pauli matrices. As this expression was derived as a first order
approximation, it is consistent to evaluate it to the same order. So,
on the RHS of Eq.(\ref{Eq-QMC22}) we replace \( M^{*} \) by \( M \)
in the denominator, we approximate \( M^{*}=M-g_{\sigma }\sigma  \)
and we use the leading approximations for the meson fields -- that is, 
\( g_{\sigma }\sigma =G_{\sigma }\rho ^{cl},\, g_{\omega }\omega 
=G_{\omega }\rho ^{cl}. \)
As the spin orbit interaction is active only at the nuclear surface,
the relevant density is rather small so these approximations are justified.

The final step is to quantize the classical Hamiltonian (\ref{Eq-QMC21})
by making the replacement 
\( \vec{P}_{i}\rightarrow -i {\nabla }_{i} \) .  
As usual we must deal with the ordering ambiguity which exists as soon as
velocity dependent interactions are present. For the spin orbit interaction
there is no ambiguity because all orderings give the same
matrix elements. The problem occurs only in the last term of the first
line of Eq.~(\ref{Eq-QMC21}). There are 2 possible hermitian orderings
when \( \vec{P}_{i} \) becomes an operator acting on the right: 
$T_{1}= ( \vec{P}^{2}_{i}\delta (\vec{R}_{ij})+\delta (\vec{R}_{ij})
\vec{P}^{2}_{i} ) /2 $
and 
\( T_{2}=\vec{P}_{i}\delta (\vec{R}_{i}-\vec{R}_{j}).\vec{P}_{i} \) . 
However, using integration by parts and the commutation rules,
one checks easily that the difference between the two orderings is
of the form \(  \nabla ^{2}_{i}\delta (\vec{R}_{ij})  \).
Such an operator is already present in \( H_{QMC} \) in the second
and third lines of Eq.~(\ref{Eq-QMC21}) and we see that choosing one
ordering or the other is equivalent to a change of the meson masses.
In practice we tried both orderings and checked that this is equivalent
to a \( 50 \) MeV change of \( m_{\sigma } \), which is not  
so well known. Since, in any case, we intend to study the sensitivity 
of our results to \( m_{\sigma } \), choosing \( T_{1} \) or \( T_{2} \)
is immaterial and for definiteness and historical reasons we adopt
the form \( T_{2}. \)

To complete our effective Hamiltonian we now include the effect of
the isovector \( \rho  \) meson, which can be done by analogy with
the \( \omega  \) meson. If we let \( b^{\alpha } \, (\alpha =1,2,3) \)
be the time component of the field and \( \tau ^{\alpha } \) the isospin
Pauli matrices, then the only changes are the replacement \( g_{\omega }\omega (\vec{R})\rightarrow g_{\omega }\omega (\vec{R})+g_{\rho }\vec{b}(\vec{R}).\vec{\tau }/2 \)
in the expression (\ref{Eq-QMC1}) for the nucleon energy and 
\( g_{\omega }\omega (\vec{R})(1-2\mu _{s})\rightarrow g_{\omega }\omega (\vec{R})(1-2\mu _{s})+g_{\rho }(1-2\mu _{v})\vec{b}(\vec{R}).\vec{\tau }/2 \)
in the expression (\ref{Eq-QMC22}) for the spin orbit interaction,
with \( g_{\rho } \) the free \( \rho  \)-nucleon coupling constant
and \( \mu _{v}=4.7 \) the nucleon isovector magnetic moment. If
we define \( G_{\rho }=g_{\rho }^{2}/m_{\rho }^{2} \) with \( m_{\rho } \)
the mass of the \( \rho  \) meson, our quantum effective Hamiltonian
finally takes the form :
\begin{eqnarray}
 &  & H_{QMC}=\sum _{i}\frac{\overleftarrow{\nabla }_{i}.
 \overrightarrow{\nabla }_{i}}{2M}+
 \frac{G_{\sigma }}{2M^{2}}\sum _{i\neq j}\overleftarrow{\nabla _{i}}
 \delta (\vec{R}_{ij}).\overrightarrow{\nabla }_{i}\nonumber \\
 &  & +\frac{1}{2}\sum _{i\neq j}\left[ \nabla ^{2}_{i}\delta (\vec{R}_{ij})\right] \left[ \frac{G_{\omega }}{m_{\omega }^{2}}-\frac{G_{\sigma }}{m_{\sigma }^{2}}+\frac{G_{\rho }}{m_{\rho }^{2}}\frac{\vec{\tau }_{i}.\vec{\tau }_{j}}{4}\right] \nonumber \\
 &  & +\frac{1}{2}\sum _{i\neq j}\delta (\vec{R}_{ij})\left[ G_{\omega }-G_{\sigma }+G_{\rho }\frac{\vec{\tau }_{i}.\vec{\tau }_{j}}{4}\right] \nonumber \\
 &  & +\frac{dG_{\sigma }^{2}}{2}\sum _{i\neq j\neq k}\delta ^{2}(ijk)-\frac{d^{2}G_{\sigma }^{3}}{2}\sum _{i\neq j\neq k\neq l}\delta ^{3}(ijkl)\nonumber \\
 &  & +\frac{i}{4M^{2}}\sum _{i\neq j}A_{ij}\overleftarrow{\nabla }_{i}
\delta (\vec{R}_{ij})\times \overrightarrow{\nabla }_{i}\, .\vec{\sigma }_{i}\, ,
\label{Eq-QMC28} 
\end{eqnarray}
with \( A_{ij}=G_{\sigma }+(2\mu _{s}-1)G_{\omega }+
(2\mu _{v}-1)G_{\rho }\vec{\tau }_{i}.\vec{\tau }_{j}/4. \) 

To fix the free parameters of the model, 
that is \( G_{\sigma },G_{\omega } \) and \( G_{\rho } \),
we have computed, using the Hamiltonian (\ref{Eq-QMC28}), the volume
and symmetry coefficients of the binding energy per nucleon of infinite
nuclear matter : \( E_{B}/A=a_{1}+a_{4}(N-Z)^{2}/A^{2} \). We have
used the experimental values \( a_{1}=-15.85{\rm MeV},\, \, a_{4}=30{\rm
MeV} \)
and the saturation condition \( \partial a_{1}/\partial \rho (\rho _{0})=0 \)
with \( \rho _{0}=0.16{\rm fm}^{-3} \). In order to avoid the proliferation
of tables we show only the results corresponding to the bag radius
\( R_{B}=0.8\, {\rm fm} \), which is realistic. We have used the physical
masses, \( m_{\omega }=782{\rm MeV},\, m_{\rho }=770{\rm MeV} \) and we allow
\( m_{\sigma } \) to take the values \( 500{\rm MeV} \) and \( 600{\rm
MeV} \),
which is a commonly accepted range. 
We get, in \( {\rm fm}^{2}, \) 
\( G_{\sigma }=12.63,G_{\omega }=9.62,G_{\rho }=9.68 \)
{}for \( m_{\sigma }=500{\rm MeV} \) and \( G_{\sigma }=11.97,G_{\omega }=8.1,G_{\rho }=6.46 \)
{}for \( m_{\sigma }=600{\rm MeV}. \) These values are larger than in the
Hartree approximation because the exchange terms tend to cancel the
direct terms of the matrix elements, thereby forcing larger couplings  
to fit the data.
\begin{table}[b]
\vspace{-0.5cm}
\caption{\label{Tab-QMB4}QMC predictions compared with the Skyrme
force.}

\begin{center}
\begin{tabular}{|c||c|c||c||c|}

\hline
&
\multicolumn{1}{c|}{QMC}&
\multicolumn{1}{c||}{QMC}&
SkIII&
QMC(N=3)\\
\cline{2-2} \cline{3-3}
\hline
\( m_{\sigma }({\rm MeV}) \)&
\( 500 \)&
\( 600 \)&
&
\( 600 \)\\
\hline
\hline
\( t_{0}({\rm MeV}\, {\rm fm}^{3}) \) &
\( -1071 \)&
\( -1082 \)&
\( -1129 \)&
\( -1047 \)\\
\hline
\( x_{0} \)&
\( 0.89 \)&
\( 0.59 \)&
\( 0.45 \)&
\( 0.61 \)\\
\hline
\( t_{3}({\rm MeV}\, {\rm fm}^{6}) \)&
\( 16620 \)&
\( 14926 \)&
\( 14000 \)&
\( 12513 \)\\
\hline
\( 3t_{1}+5t_{2}({\rm MeV}\, {\rm fm}^{5}) \)&
\( 192 \)&
\( 475 \)&
\( 710 \)&
\( 451 \)\\
\hline
\( 5t_{2}-9t_{1}({\rm MeV}\, {\rm fm}^{5}) \)&
\( -7622 \)&
\( -4330 \)&
\( -4030 \)&
\( -4036 \)\\
\hline
\( W_{0}({\rm MeV}\, {\rm fm}^{5}) \)&
\( 118 \)&
\( 97 \)&
\( 120 \)&
\( 91 \)\\
\hline
\( K({\rm MeV}) \)&
\( 327 \)&
\( 327 \)&
\( 355 \)&
\( 364 \)\\
\hline
\end{tabular}\end{center}
\end{table}

As a practical test of the capacity of our model 
to interpret a large body of nuclear
data, we compare it with the effective Skyrme interaction. Since, in
our formulation, the medium effects are summarized in the 3 and 4
body forces, we consider Skyrme forces of the same type, that is without density dependent interactions. They are
defined by a potential energy of the form :
\begin{eqnarray}
 &  & V=t_{3}\sum _{i<j<k}\delta (\vec{R}_{ij})\delta (\vec{R}_{jk})+t_{0}\sum _{i<j}(1+x_{0}P_{\sigma })\delta (\vec{R}_{ij})\nonumber \\
&  & +\frac{1}{4}t_{2}\, \overleftarrow{\nabla }_{ij}.\delta (\vec{R}_{ij})
\overrightarrow{\nabla }_{ij}-
\frac{1}{8}t_{1}\left[ \delta (\vec{R}_{ij})\overrightarrow{\nabla }_{ij}^{2}+
\overleftarrow{\nabla }_{ij}^{2}\delta (\vec{R}_{ij})\right] \nonumber \\
&  & +\frac{i}{4}W_{0}(\vec{\sigma }_{i}+\vec{\sigma }_{j}).
\overleftarrow{\nabla }_{ij}
\times \delta (\vec{R}_{ij})\overrightarrow{\nabla }_{ij},
\label{Eq-QMC53} 
\end{eqnarray}
with \( \nabla _{ij}=\nabla _{i}-\nabla _{j} \). There is no 4-body
force in Eq.~(\ref{Eq-QMC53}) and we shall show its possible impact by
setting its strength equal to zero in \( H_{QMC} \). Since the spin
exchange operator, \( P_{\sigma } \), in \( V \) multiplies a contact
interaction, its action on an antisymmetric state it is equivalent
to minus the isospin exchange operator. Comparison of Eq.~(\ref{Eq-QMC53})
with the QMC Hamiltonian (\ref{Eq-QMC28}) allows one to identify
\begin{equation}
t_{0}=-G_{\sigma }+G_{\omega }-\frac{G_{\rho }}{4},\, \, \, 
t_{3}=3dG_{\sigma }^{2},\, \, \, x_{0}=-\frac{G_{\rho }}{2t_{0}} \, .\label{Eq-QMC100}\end{equation} 

{}For the other parameters we cannot make a direct identification because,
as our effective Hamiltonian is derived in the rest frame of the nucleus,
its momentum dependent pieces violate Galilean invariance. This is
irrelevant since it is devised for variational calculations where the
trial state also violates Galilean invariance but to make the identification
we need to compare the respective Hartree-Fock Hamiltonians rather
than the interactions themselves. To this end we make some simplifying
assumptions which do not significantly 
damage the physics but avoid unnecessary  
technical complications. First, we restrict our considerations to doubly
closed shell nuclei with \( N=Z. \) Second we assume that one can
neglect the difference between the radial wave functions of the single
particle states with \( j=l+1/2 \) and \( j=l-1/2 \) . This amounts
to treating the spin orbit interaction to first order, which is sufficient
for our purposes. By comparing the Hartree-Fock Hamiltonian obtained
from \( H_{QMC} \) and that of Ref.~\cite{Vauth72} corresponding
to the Skyrme force , we obtain the relations :
\begin{eqnarray}
 &  & 3t_{1}+5t_{2}=\frac{8G_{\sigma }}{M^{2}}+4\left( \frac{G_{\omega }}{m_{\omega }^{2}}-\frac{G_{\sigma }}{m_{\sigma }^{2}}\right)
  +3\frac{G_{\rho }}{m_{\rho }^{2}}\, ,\label{EqQMC110}\\
 &  & 5t_{2}-9t_{1}=\frac{2G_{\sigma }}{M^{2}}+28\left( \frac{G_{\omega }}{m_{\omega }^{2}}-\frac{G_{\sigma }}{m_{\sigma
 }^{2}}\right) -3\frac{G_{\rho }}{m_{\rho }^{2}}\, ,\\
 &  & W_{0}=\frac{1}{12M^{2}}\left( 5G_{\sigma }+5(2\mu _{s}-1)G_{\omega }+\frac{3}{4}(2\mu _{v}-1)G_{\rho }\right)\, .\nonumber\\
 &  & 
\end{eqnarray}

In Table~\ref{Tab-QMB4} we compare our results with the parameters
of the force SkIII \cite{Fried86}, which is considered a good representative
of density independent effective interactions. We postpone a more
extensive comparison to future work. 
Instead of  $t_1,t_2$, we show the combinations \( 3t_{1}+5t_{2} \),
which controls the effective mass, and \( 5t_{2}-9t_{1} \), which controls
the shape of the nuclear surface\cite{Vauth72}. 
Even bearing in mind that, since we use
the parameters \( (a_{1},a_{4},\rho _{0}) \) as input, not all the
numbers in Table~\ref{Tab-QMB4} are predictions, we see that the
level of agreement with SkIII, for \( m_{\sigma }=600{\rm MeV} \),   
is still impressive. An important point is that the spin orbit strength, 
\( W_{0} \), comes out with approximately the correct value, independent  
of the \( \sigma  \) mass. The last column (N=3) shows our results when
we switch off the 4-body force. The main change is a decrease of the
predicted 3 body force. Clearly this mocks up the effect of the attractive
4-body force which may then appear less important. However, this
is misleading if we look at the compressibility of nuclear matter,
\( K=9\rho ^{2}\partial ^{2}a_{1}/\partial \rho ^{2} \)
(last line of Table~\ref{Tab-QMB4}), which decreases by as much
as \( 37{\rm MeV} \) when we restore this 4-body force. The value we
find, \( K=327{\rm MeV} \), is still a little too large with respect to the
experimental range (\( 200\div 300{\rm MeV} \)), but several
simplifications made for this presentation can be eliminated in future
work. Moreover we have not yet included 
the long-range force of the pion, which will reduce \( K \) by about
\( 20{\rm MeV} \) according to a preliminary calculation. This too will be
investigated in future work\cite{InProgress}.

In summary, we have demonstrated a remarkable agreement between the
phenomenologically successful Skyrme force, SkIII,
and the effective interaction
corresponding to the quark meson coupling model -- a result which
suggests that the response of nucleon internal structure to the nuclear
medium does indeed play a vital in nuclear structure.

We acknowledge the support of the French CEA as well as 
the CSSM, where this investigation began.
This work is supported by the Australian Research
Council.
\end{document}